\documentclass[useAMS,usenatbib,usegraphicx,uselongtable]{mn2e}
\usepackage[varg]{txfonts}

\usepackage{longtable}
\usepackage{lscape}
\usepackage{graphicx}
\usepackage{rotating}

\title[Brown dwarf disks at 50 Myr]{A Herschel PACS survey of brown dwarfs in IC 2391: Limits on primordial and debris disk fractions}
\author[Riaz \& Kennedy]
{B. Riaz$^{1}$ \& G. M. Kennedy$^{2}$ \\
$^{1}$Centre for Astrophysics Research, Science \& Technology Research Institute, University of Hertfordshire, Hatfield, AL10 9AB, UK \\
$^{2}$Institute of Astronomy, University of Cambridge, Madingley Road, Cambridge CB3 0HA, UK  }

\begin{document}

\date{}

\pagerange{\pageref{firstpage}--\pageref{lastpage}} \pubyear{2014}

\maketitle

\label{firstpage}

\begin{abstract}

We present results from a Herschel PACS survey of 8 brown dwarfs in the IC 2391 cluster. Our aim was to determine the brown dwarf disk fraction at ages of $\sim$40--50 Myr. None of the 8 brown dwarfs observed were detected in the PACS 70 or 160$\mu$m bands. We have determined the detection limits of our survey using the 1-$\sigma$ flux upper limits in the PACS far-infrared and the WISE mid-infrared bands. The sensitivity of our observations would only allow for the detection of debris disks with exceptionally large fractional luminosities ($\geq$1\%). Considering that only the most extreme and rare debris disks have such high fractional luminosities, it can be hypothesized that Vega-like debris disks, as observed around $\sim$30\% of low-mass stars at similar ages, could exist around the targeted IC 2391 brown dwarfs. Most primordial disks similar to the ones observed for the younger 1--10 Myr brown dwarfs would be within the detection sensitivities of our survey, and could have been detected. The non-detection for all targets then suggests that brown dwarf disks have transitioned to the debris phase by $\sim$40--50 Myr ages. We also present the sensitivity limits for detecting brown dwarf disks with future SPICA observations. 

\end{abstract}

\begin{keywords}
circumstellar matter -- planetary systems: formation -- stars: low-mass, brown dwarfs -- open clusters and associations: individuals (IC 2391)
\end{keywords}

\section{Introduction}
\label{intro}

Low-mass pre-main sequence stars are formed with substantial rotating disks around them. These optically thick primordial disks are massive and have an initial gas:dust ratio similar to the interstellar ratio. As the system evolves, the primordial disk gradually dissipates, resulting in a decrease in the total disk mass. The dissipation may occur through various processes. Much of the disk may be accreted onto the central star; some material may be driven off by stellar winds \citep[e.g.,][]{gull, muz, calvet}. The dissipation may also occur through the coagulation of the smaller particles to form larger planetesimals, which may eventually lead to planet formation \citep[e.g.,][]{dullemond, quillen, rice, setiawan}. The collisions among the planetesimals produce dust grains that form the secondary debris disks. These are gas-poor disks, comprised of dust which is continually regenerated as the larger and more massive planetesimals trigger shattering collisions amongst the smaller bodies. Dissipation in the debris disks occurs as the mass reservoir of large planetesimals, which supplies the observed dust, is depleted \citep[e.g.,][]{wyatt}, whereas the smallest ($<$1$\mu$m) dust is removed by radiation and stellar wind forces \citep[e.g.,][]{burns, plavchan}. 


Recent near- and mid-infrared surveys of clusters and associations spanning a range in ages and masses have revealed extensive information on the disk frequencies and lifetimes. The main finding from these surveys is that optically thick primordial disks are extensively dissipated by an age of $\sim$10 Myr  \citep[e.g.,][]{strom, haisch, young, hillen}. It is also at this critical age of $\sim$10 Myr that a rise in the debris disk frequency is observed. Debris disks are commonly detected at $\sim$10-30 Myr as they are bright and have not had much time to decay yet. Thereafter, the frequency declines roughly as 1/t over a time scale of a few hundred Myr \citep[e.g.,][]{siegler, rieke, c09}. 

Also notable is the dependence of the decay timescales on the spectral type of the central source. As discussed in \citet{r12}, the primordial disk lifetime appears prolonged in very low-mass stars and brown dwarfs compared to the higher mass stars, such that nearly all brown dwarf disks at an age of $\sim$10 Myr are in the primordial stage, compared to a $\sim$10\% fraction for low-mass stars (0.1--1$M_{\sun}$) and a null fraction for $>$1$M_{\sun}$ stars. For the debris disks, the inner disk decay timescales, as probed by 24$\mu$m dust emission, appear shorter for the FGKM stars compared to the more massive B- and A-type stars \citep[e.g.,][]{siegler, beichman}. The rapid inside out evacuation of disk material suggests that processes such as planet formation are completed inside $\sim$1 AU within 30--50 Myr for the low-mass stars, resulting in very few 24$\mu$m disk detections at these later ages \citep[e.g.,][]{gorl, spezzi}. Other processes such as, Poynting-Robertson (P-R) drag, stellar wind drag, or planet-dust dynamical interactions may also be responsible for shortening the dust grain removal timescales \citep[e.g.,][]{plavchan}. It should be noted that 24$\mu$m observations of nearby stars are more sensitive, which explains why a higher fraction ($\sim$5\%; Trilling et al. 2008) of these stars are seen to have 24$\mu$m excesses, compared to observations of young clusters such as IC 2391 and NGC 2547. On the other hand, the differences in the disk frequency trends could be due to a luminosity effect rather than a mass-dependent effect, as the more massive stars heat up dust at larger annuli to the levels detectable at mid- and far-infrared wavelengths. The low stellar luminosities could make even a 10AU disk so cold that it can only be detected at far-infrared wavelengths, as observed to be the case for the disk around Fomalhaut C, which is only detected at $>$100$\mu$m wavelengths (Kennedy et al. 2014). 

The search for debris disks around brown dwarfs is a field which has not been explored to great lengths yet. Previous surveys on brown dwarf disks based on mid- and far-infrared observations have all focused on ages of $<$10 Myr, at which brown dwarf disks are still found to be in the primordial phase (Riaz \& Gizis 2008; Riaz et al. 2012; Harvey et al. 2012). We have conducted a photometric survey of brown dwarfs in the $\sim$40--50 Myr old cluster IC 2391, using the ESA {\it Herschel} Space Observatory \citep{pilbratt} Photodetector Array Camera and Spectrometer (PACS) \citep{pog} instrument (70 and 160$\mu$m), to search for cold dust emission that would indicate the presence of debris disks. Considering the trends noted above, no clear transition from a primordial to a debris phase is observed for the brown dwarf disks even at $\sim$10 Myr ages, unlike the higher mass stars. Our main scientific goal is to map the evolution of brown dwarf disks over a wider age range, and to determine if the onset of the debris phase occurs at a later age for the sub-stellar sources, compared to the earlier-type stars. 


\section{Observations and Data Analysis}
\label{obs}

We searched for nearby pre-main-sequence clusters with an estimated age of $\sim$30--50 Myr, and which contain spectroscopically confirmed sub-stellar members with masses below 0.08$M_{\sun}$. At $\sim$30--50 Myr, this sub-stellar mass limit corresponds to a T$_{eff}$ of less than $\sim$2900 K and a spectral type $\geq$M6 \citep[e.g.,][]{baraffe}. The IC 2391 cluster is therefore of interest. It has an age of 50 Myr estimated via the lithium method \citep{b04} and 40 Myr from main-sequence fitting (Platais et al. 2007), and lies at a distance of 145$\pm$2.5 pc \citep{van}. There are currently about 10 spectroscopically confirmed brown dwarfs in this cluster with spectral type $\geq$M6 and masses below 0.07 $M_{\sun}$ \citep{b09, b04}. We obtained PACS 70 and 160$\mu$m photometric observations under {\it Herschel} Cycle 2 (PID: OT2\_briaz\_4) for 8 brown dwarfs in IC 2391 (Table~\ref{phot}). All observations were made using the mini-scan map mode. Two cross-scans were obtained at orientation angles of 77$\degr$ and 110$\degr$, with a scan leg length of 2.5$\arcmin$, cross-scan step of 4$\arcsec$, and 10 scan legs for each scan. The medium scan speed (20$\arcsec$/s) was used as it provides optimum sensitivity. The 1-$\sigma$ rms noise level using 6 repeats of this setup was expected to be $\sim$1 mJy at 70$\mu$m, using the HSPOT tool. 

\begin{figure*}
\centering
  \includegraphics[width=14cm, angle=0]{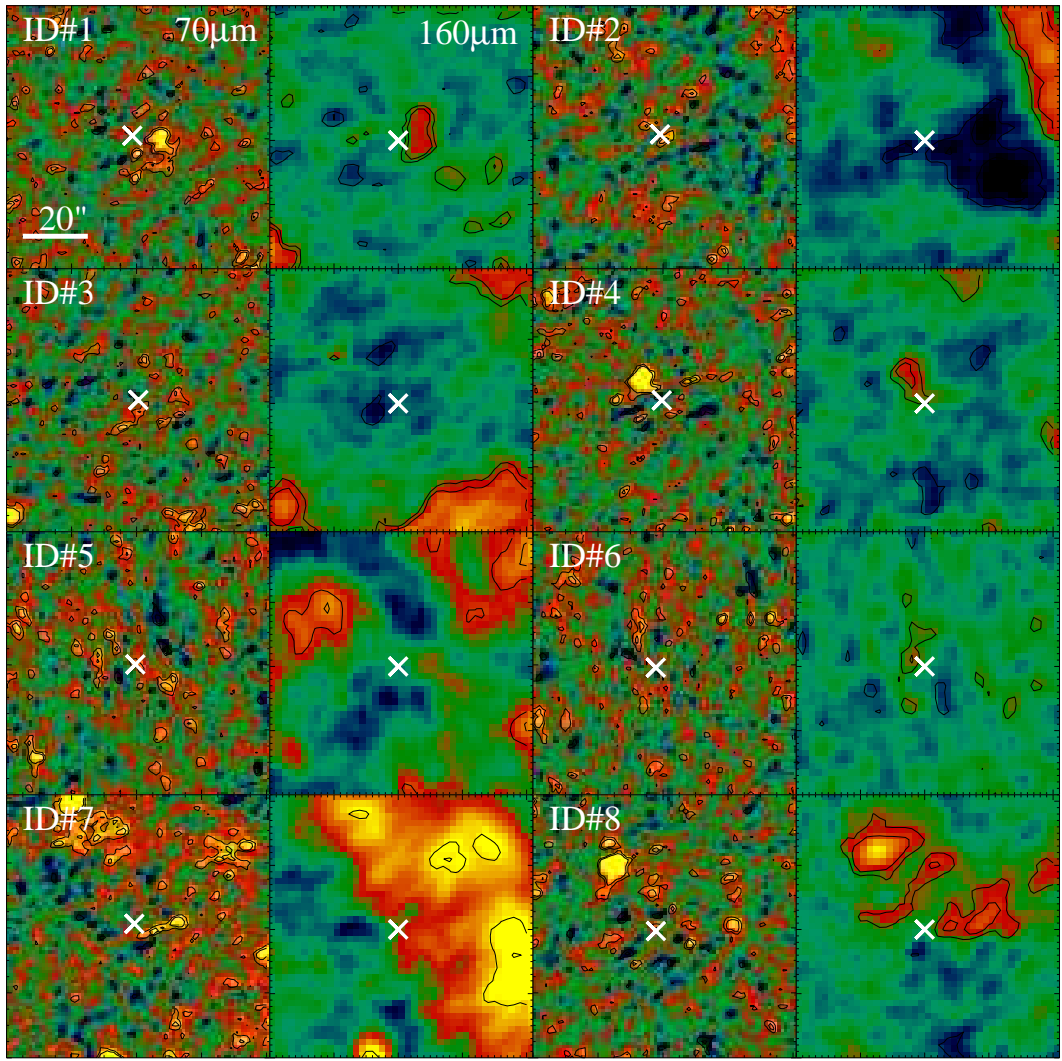}  
 \caption{PACS 70 and 160$\mu$m images for the eight targets. White crosses mark the target positions, which are close to  the centre of each image. Each image is 80\arcsec square, and the colour scale is a linear stretch from $-5$ to $8 \times 10^{-5}$ Jy pixel $^{-1}$ at 70 $\mu$m, and $-4$ to $8 \times 10^{-4}$ Jy pixel $^{-1}$ at 160 $\mu$m. Contours are shown at -3, -2, 2, and 3 times the pixel to pixel RMS for each image. North is up, East is to the left.  }
  \label{images}
\end{figure*}

\begin{figure}
\centering
  \includegraphics[width=9cm, angle=0]{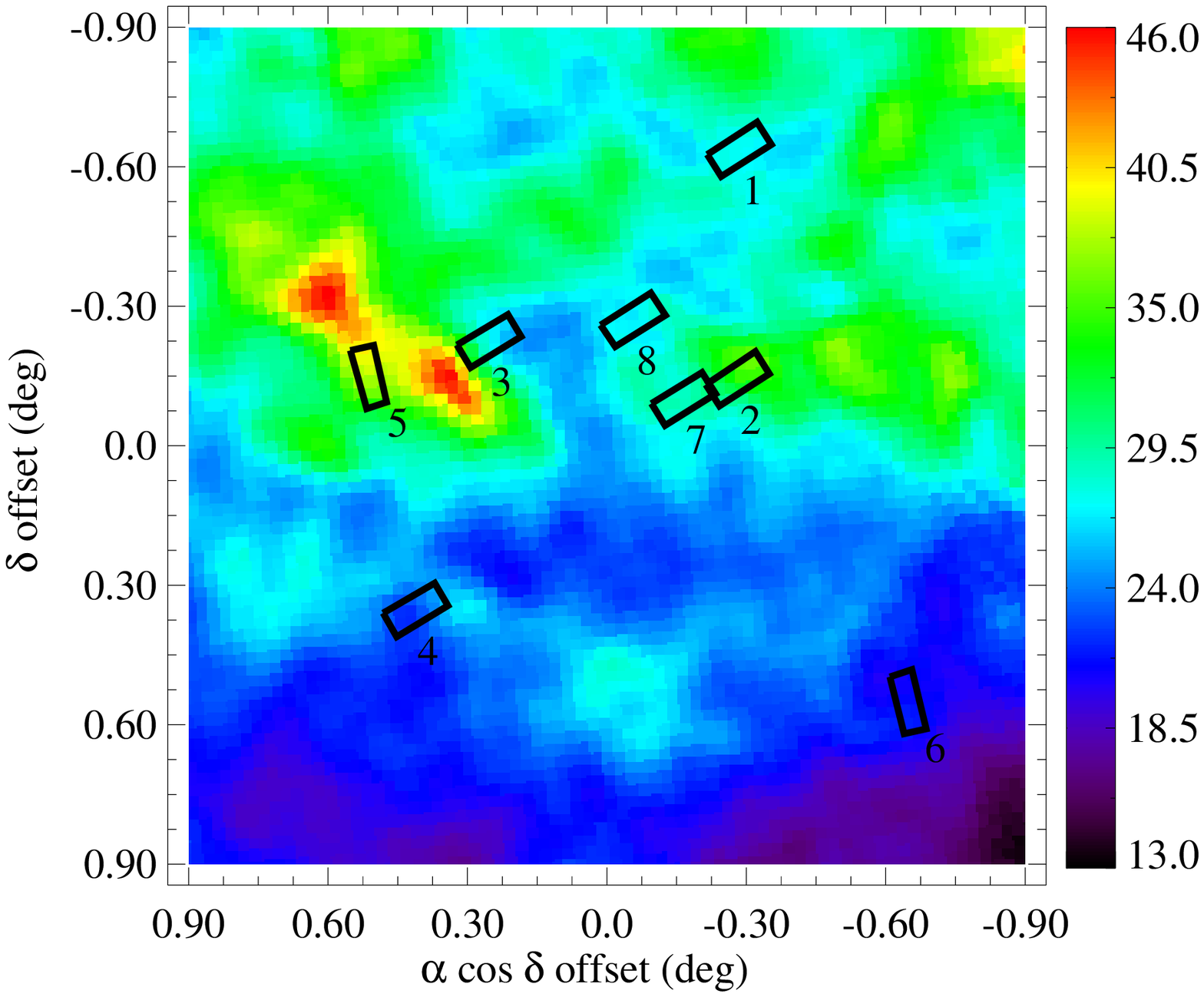}   
 \caption{A comparison of the PACS image locations with the IRAS 100$\mu$m noise levels \citep{miville}. The scale on the right is in MJy sr$^{-1}$. The centre location (0,0) is at RA = 8h 40m 39s, Dec = -52d 53m 18s. The labels are the same as the IDs in Table~\ref{phot}. }
  \label{IRAS}
\end{figure}

Data analysis was performed on the `Level 2' data products processed using the {\it Herschel} Interactive Processing Environment (HIPE version 11.0.1). We created a mosaic of the two scan maps obtained at different orientation angles. The 160$\mu$m images do not cover exactly the same sky area as the 70$\mu$m images, even though these were observed simultaneously; one of the red band arrays expired in the last few months of {\it Herschel} operation, which is presumably the reason for the difference in the coverage area. None of the targets were detected in any of the PACS bands at the expected source location in the maps. The objects with ID \# 1 and 4 have a point source located about 10$\arcsec$ away in each case (Fig.~\ref{images}), but considering that the {\it Herschel} pointing was good to about 1$\arcsec$, these sources are unrelated to the targets in question. There is some scattered emission observed close to the target location for objects with ID \# 3, 6, and 8, which cannot be confirmed as a point-like source detection. We measured the 1-$\sigma$ noise in the maps by placing a large number of apertures at random locations in each map, and then taking the standard deviation of these measurements. The aperture sizes were set to 4$\arcsec$ and 8$\arcsec$, which are the optimal signal-to-noise ratio (SNR) sizes at 70 and 160$\mu$m. The 1-$\sigma$ noise levels thus derived are listed in Table~\ref{phot}. The noise measurements are not as similar to one another at 160$\mu$m as expected given the same integration times. However, putting all the images together on the sky and comparing with the IRAS 100$\mu$m levels (Fig.~\ref{IRAS}) seems to give sensible results when compared with the derived noise levels. The 160$\mu$m noise is quite a bit higher than at 70$\mu$m. This is in part due to instrument sensitivities, but is also related to the loss of some of the red array. Looking at the individual images, there is some background structure that will have also contributed (Fig.~\ref{IRAS}). The targets with ID \# 2, 5, and 7 lie along a brighter region and also have higher noise, whereas, ID \# 6, 1, and 4 have the flattest looking maps and these have the lowest 160$\mu$m noise. The 70$\mu$m maps thus appear sky noise limited, while the 160$\mu$m maps are background limited, with perhaps ID \# 6 being sky noise limited at 160$\mu$m. 

\begin{figure}
\centering
  \includegraphics[width=4cm, angle=0]{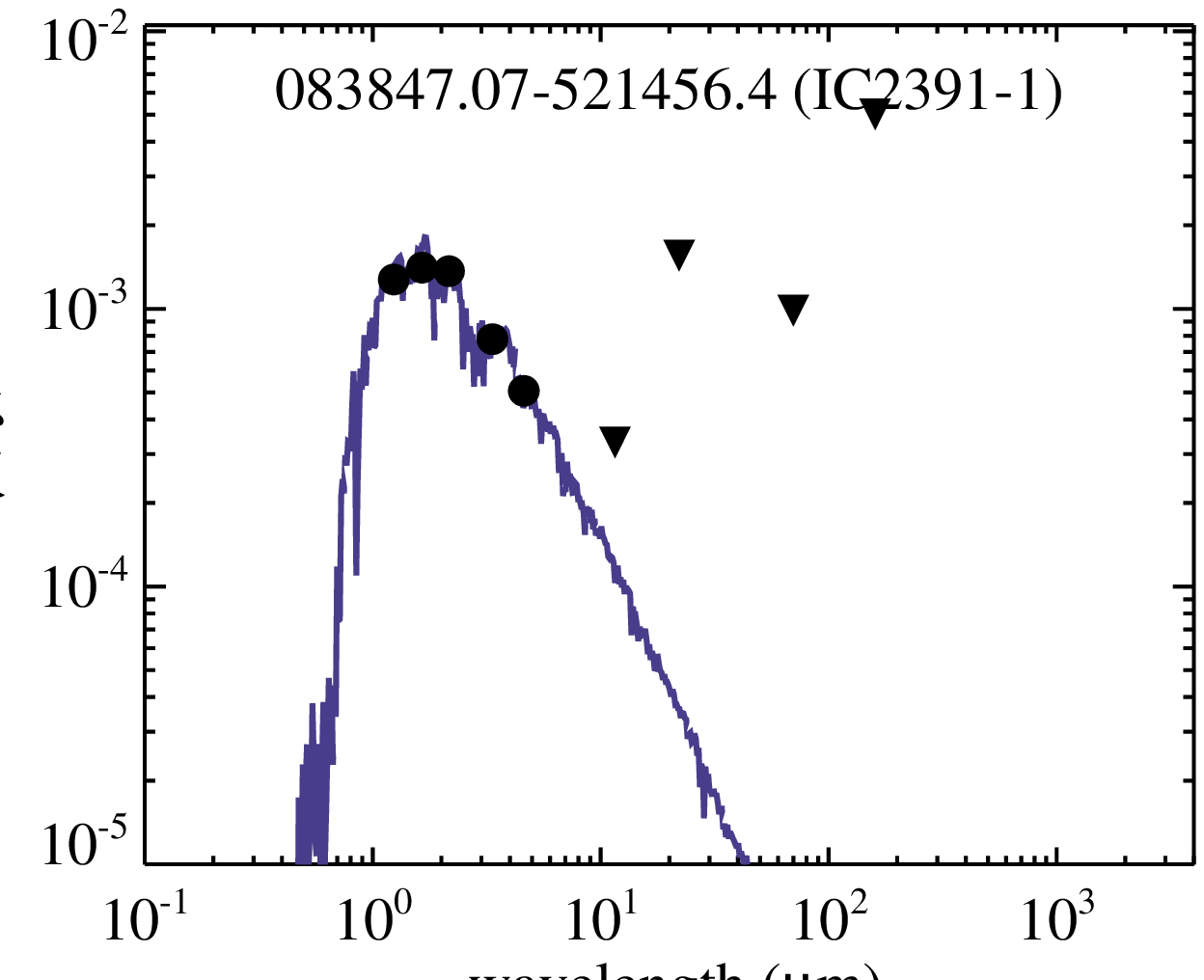}   
  \includegraphics[width=4cm, angle=0]{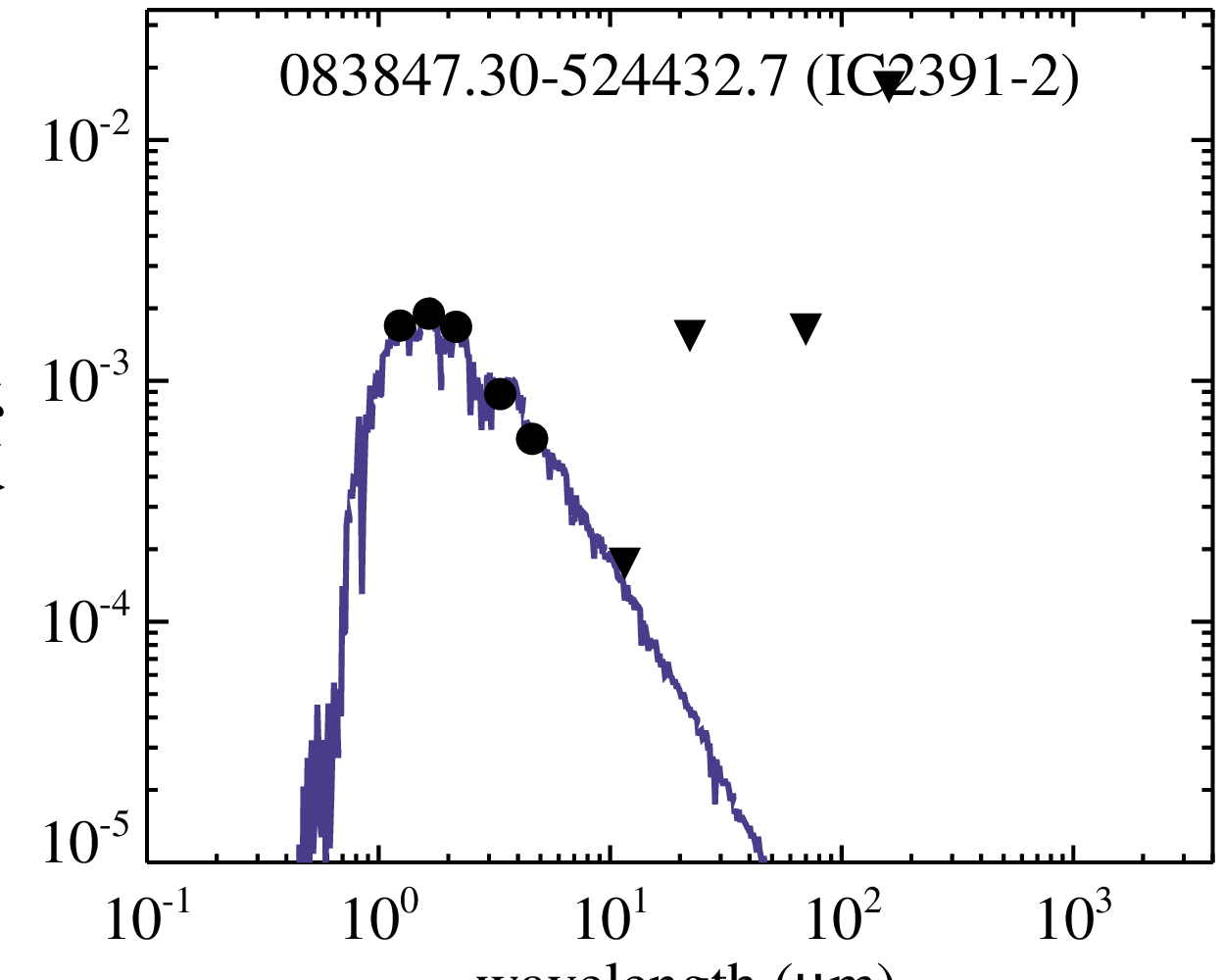}   \\
  \includegraphics[width=4cm, angle=0]{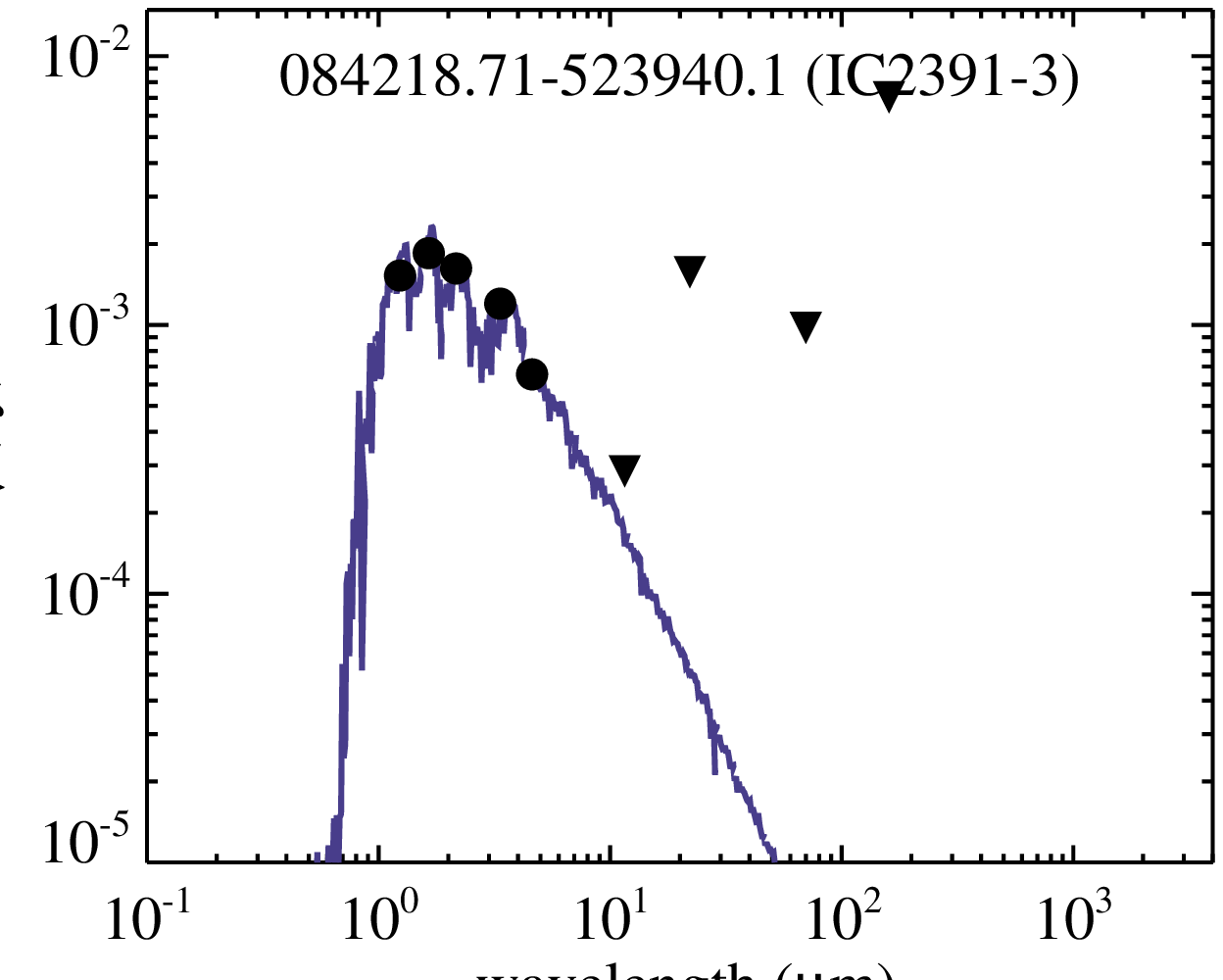}   
  \includegraphics[width=4cm, angle=0]{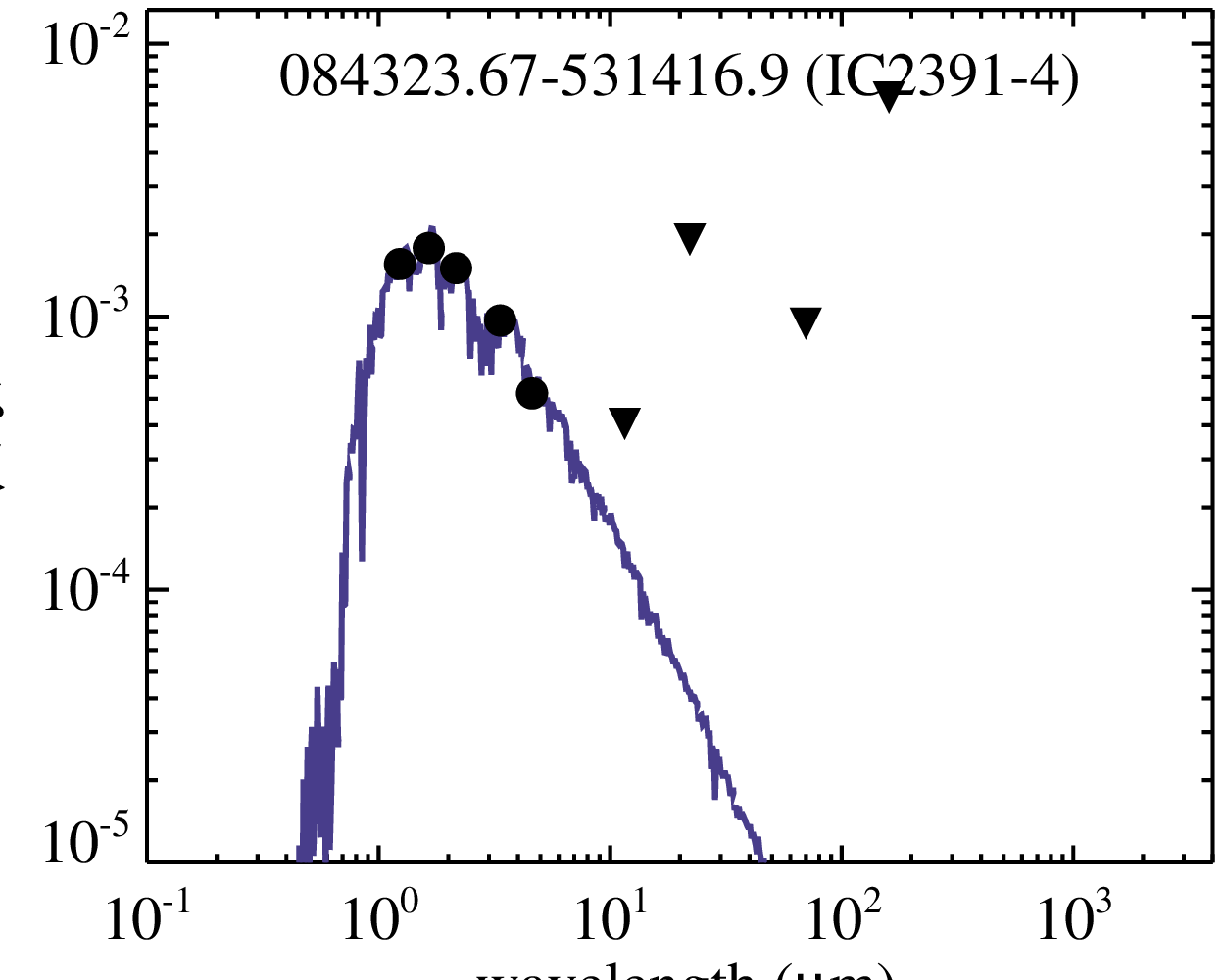}   \\
  \includegraphics[width=4cm, angle=0]{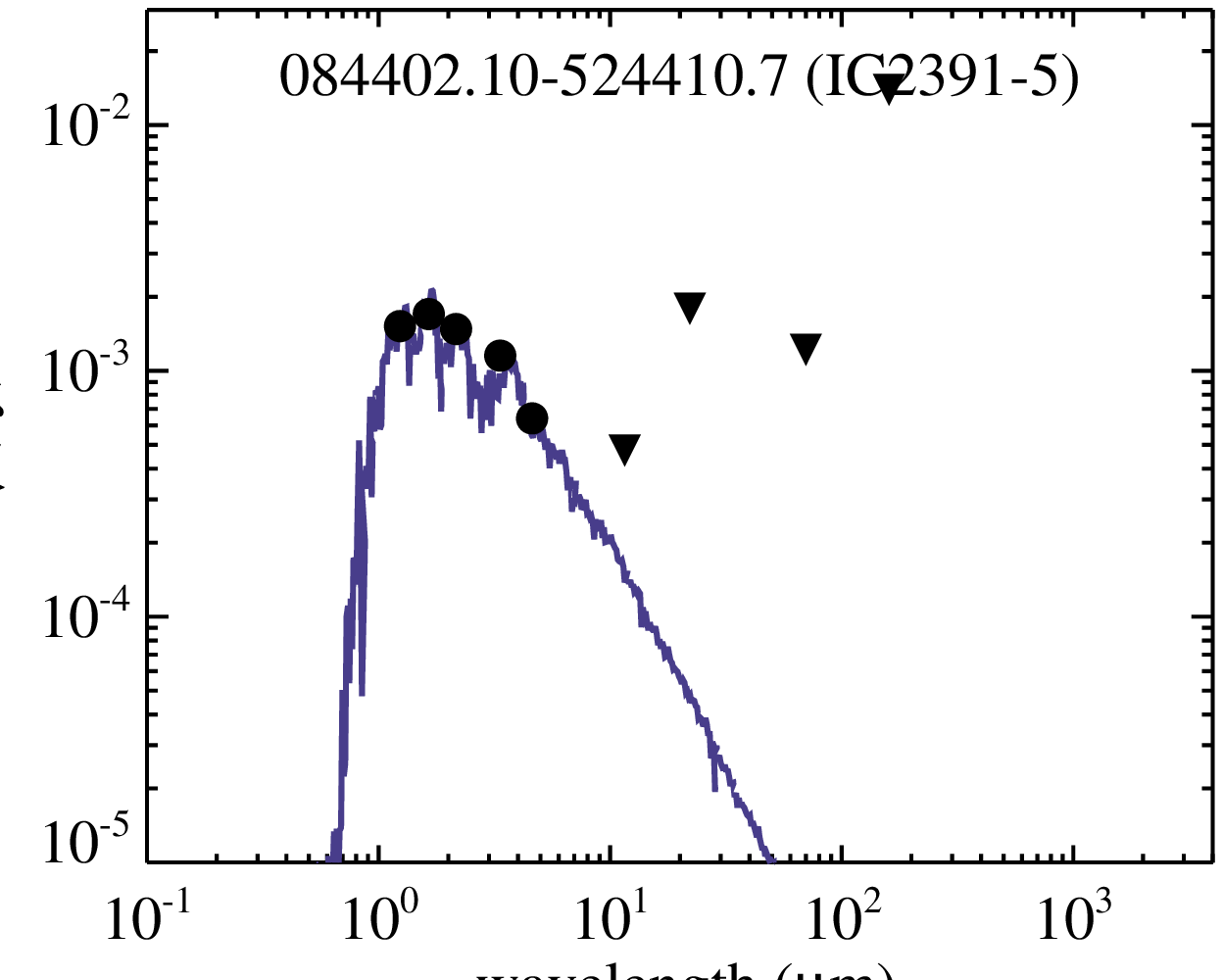}   
  \includegraphics[width=4cm, angle=0]{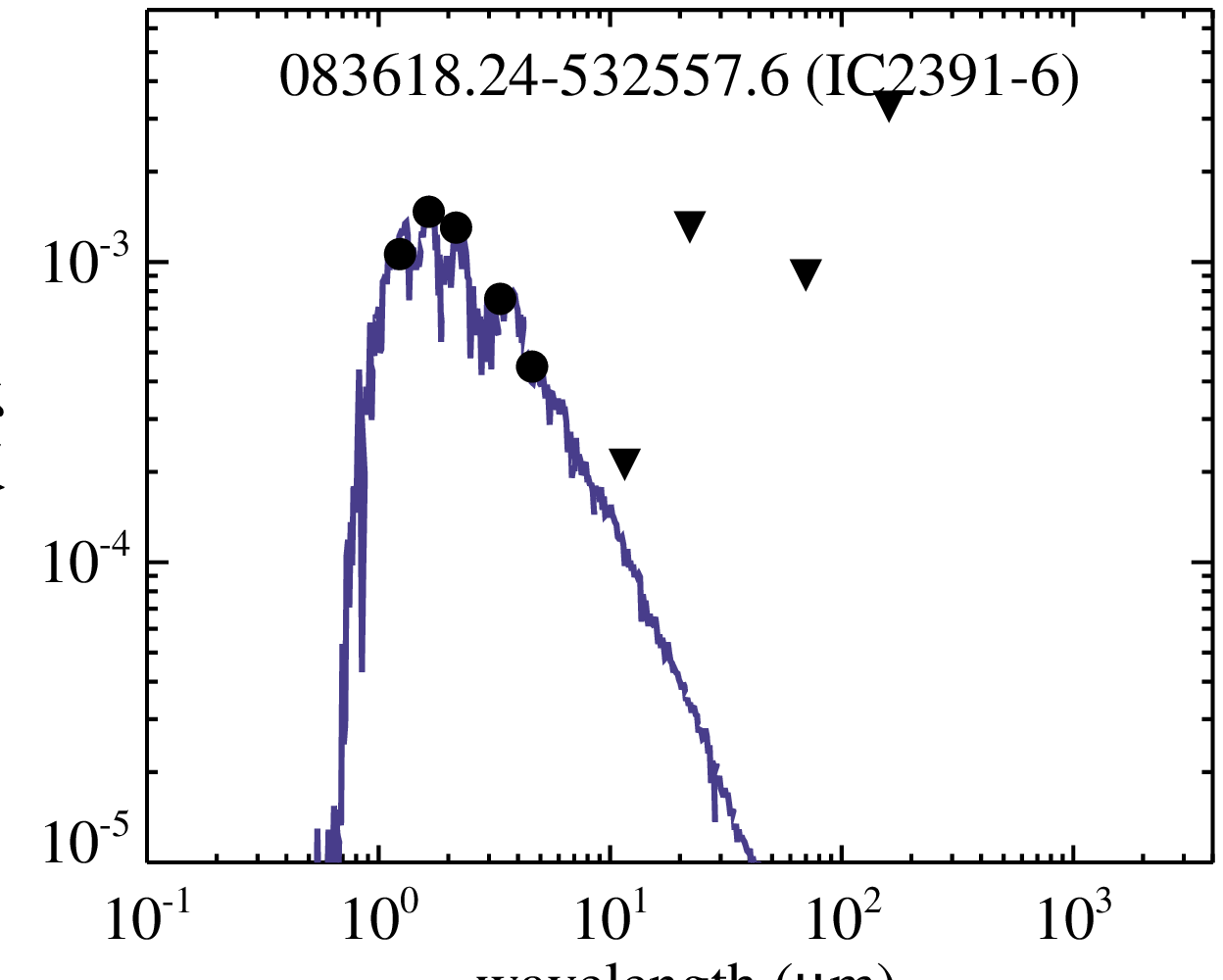}   \\
  \includegraphics[width=4cm, angle=0]{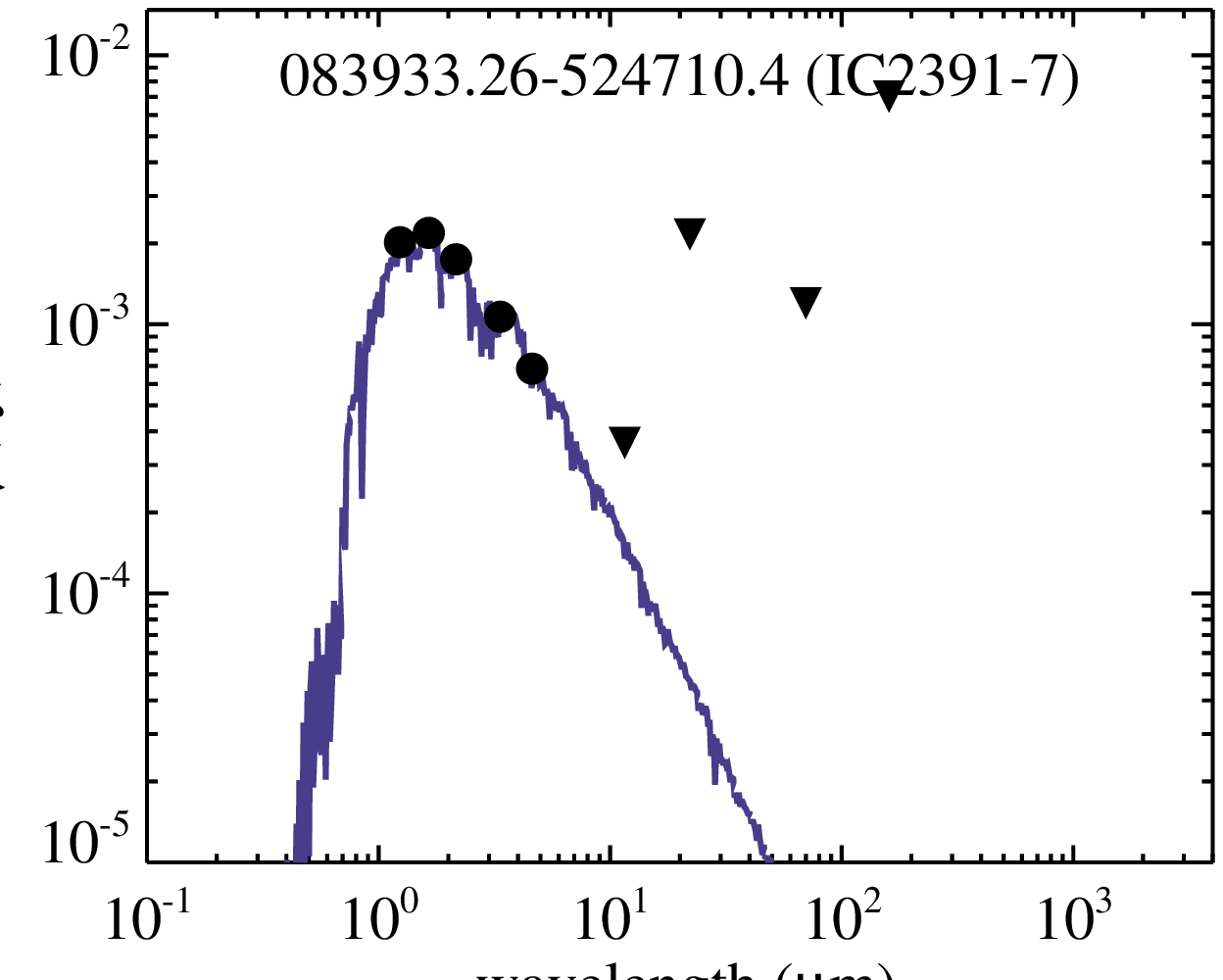}   
  \includegraphics[width=4cm, angle=0]{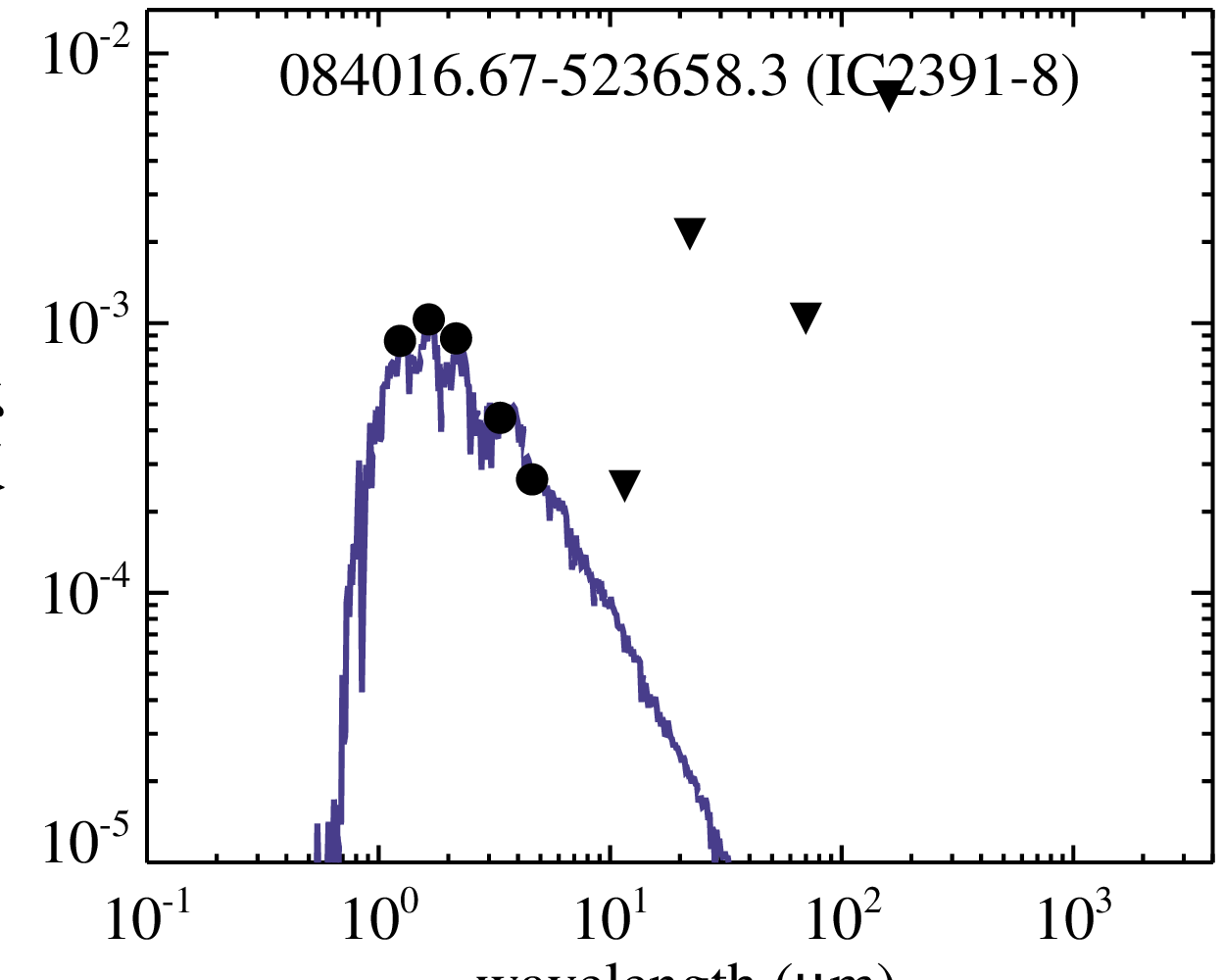}   \\ 
 \caption{SEDs for the targets. Solid line is the BT Settl model fit to the photosphere. The filled circles and triangles are the observed photometry and upper limits, respectively. The photometry for the target ID \# 5 was found in the AllWISE Reject Table, because it is affected by a diffraction spike from a nearby star. }
  \label{SED}
\end{figure}

We also checked for matches in the Wide-field Infrared Survey Telescope (WISE) \citep{wise} database. We used the ALLWISE Source Catalog, which provides better sensitivities and depth-of-coverage than the WISE All-Sky Source Catalog. Out of the 8 targets, 7 objects have WISE counterparts within 10$\arcsec$ of the target position, while one target (ID\#5) is listed in the ALLWISE Reject Catalog due to contamination by a diffraction spike from a nearby source. All of the 7 objects with WISE counterparts are detected in the 3.4 and 4.6$\mu$m bands at a SNR $>$ 20 (Table~\ref{phot}). None are convincingly detected in the 12 and 22$\mu$m bands, as indicated by the low SNR and by checking the individual WISE images. We have considered the 12 and 22$\mu$m photometry as upper limits for all sources. The spectral energy distributions (SEDs) for the targets combining all photometry are shown in Fig.~\ref{SED}. We have used the BT Settl models \citep{allard} to fit the photospheric emission. The normalisation of the atmosphere fits is to the ($J$-[4.6]) photometry via least squares fitting, with $T_{eff}$ fixed. The filled circles and triangles are the observed photometry and upper limits, respectively. The upper limits in the WISE 12 and 22$\mu$m bands and the PACS bands appear in excess to the photospheric emission for all targets. 


\section{Results and Discussion}
\label{results}

\subsection{Fractional Debris Disk Luminosities}

\begin{figure}
\centering
  \includegraphics[width=8.5cm, angle=0]{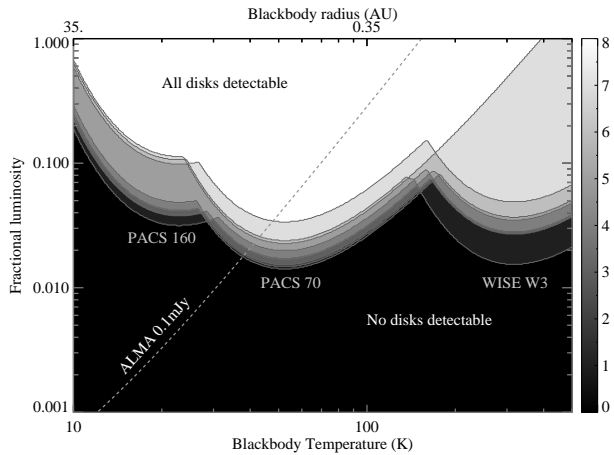} 
 \caption{Sensitivity to blackbody emission around survey targets (3$\sigma$). The top and bottom axes show the corresponding radius and blackbody temperature, respectively. Each of the eight lines shows the sensitivity for a specific target (IDs marked on the right axis), with different minima corresponding to different instruments of PACS 160 \& 70$\mu$m and WISE 12$\mu$m (W3) (from left to right). Each contour level shows the number of brown dwarfs around which disks at specific locations in the parameter space could have been detected. Also shown is the expected sensitivity for an ALMA observation with a 3$\sigma$ noise level of 0.1mJy. }
  \label{lum}
\end{figure}

The non-detection of cold debris dust could be due to the poor sensitivities of our survey. To probe further into how bright a debris disk around the targeted brown dwarfs would need to be in order to be detectable, we have converted the detection limits (or the 1-$\sigma$ upper limits) into the fractional disk luminosity, $L_{disk}$/$L_{*}$, for each source, as shown in Fig.~\ref{lum}. The contours in the plot show how bright a disk with a specific temperature and fractional luminosity could have been detected around how many of the targeted brown dwarfs, given the upper limits obtained from the observations. We assume that the gaseous material has dispersed, therefore the fractional luminosities correspond to dust emission only. We assume blackbody emission from a disk at some specific radius, but the limits are otherwise entirely model independent. The bottom axis shows the disk blackbody temperature, and the top axis shows the corresponding radius where the dust material is located, assuming $L_\star=0.002L_\odot$. Each individual curve shows the combined limits for each brown dwarf, with different troughs corresponding to PACS 160 and 70$\mu$m, and the WISE 12$\mu$m bands (from left to right, as labelled). The WISE 22$\mu$m curves are present, but barely visible since the PACS and WISE 12$\mu$m data give better sensitivity at temperatures of around 200 K. Each minimum corresponds to the blackbody peak at one of the observed wavelengths (i.e. PACS 70$\mu$m is most sensitive to 50 K emission), showing that observations at a range of wavelengths are needed to constrain emission at a range of disk radii. 

Combining all contours shows how many of the eight observations could have detected a disk at any point in the given parameter space. No disks with fractional luminosities lower than about 1\% could have been detected for any target, whereas disks with fractional luminosities greater than about 10\% could have been detected for most targets at most disk temperatures. Also plotted is the expected sensitivity for an ALMA observation with a 3$\sigma$ noise level of 0.1mJy, showing that ALMA is sensitive to brown dwarf debris disks with fractional luminosities $\gtrsim$0.1\% that lie at a few tens of AU.

For low-mass stars with $T_{eff}$ of 3000--5000 K and at ages of $\sim$10--400 Myr, the observed fractional dust luminosities range between 10$^{-5}$ and 10$^{-3}$ \citep[e.g.,][]{low, l06, l12, plavchan}. Among these is the well-known M dwarf debris disk system of AU Mic ($\sim$12 Myr old), with a fractional luminosity of $\sim$6$\times$10$^{-4}$ \citep{liu}. For nearby (within $\sim$20 pc) FGK type fields stars ($\sim$0.1--10 Gyr) observed under the {\it Hershcel} DUNES survey, the fractional dust luminosities are estimated to be between $\sim$7$\times$10$^{-7}$ and 3$\times$10$^{-4}$ \citep{eiroa}. Among the M dwarfs observed by \citet{siegler} in IC 2391, even the brightest debris disk may just be approaching a $\sim$1\% fractional luminosity, depending strongly on the assumed dust temperature. At younger ages, the $\sim$10 Myr old M dwarf debris disks TWA 7 and TWA 13AB have fractional luminosities of 10$^{-3}$--10$^{-4}$ \citep{low}. The faintness of late-type stars means that detecting infrared excesses becomes increasingly difficult, and detections are therefore biased towards the brightest disks \citep[e.g.,][]{plavchan}. The PACS observations are therefore not sensitive enough to detect debris disks around IC 2391 brown dwarfs at fractional luminosities similar to those observed among late-type stars, which are at the same or even younger ages.

\subsection{Disk Fractions versus Stellar Mass and Age}

\begin{figure*}
\centering
  \includegraphics[width=14cm, angle=0]{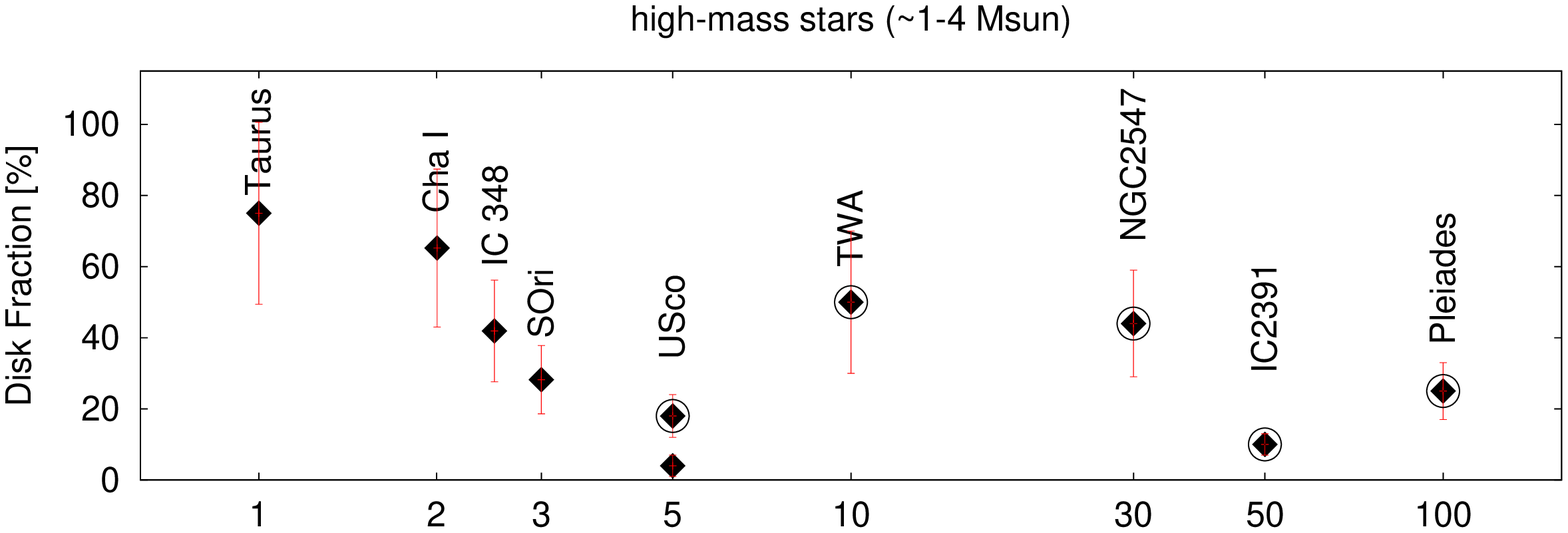}
  \includegraphics[width=14cm, angle=0]{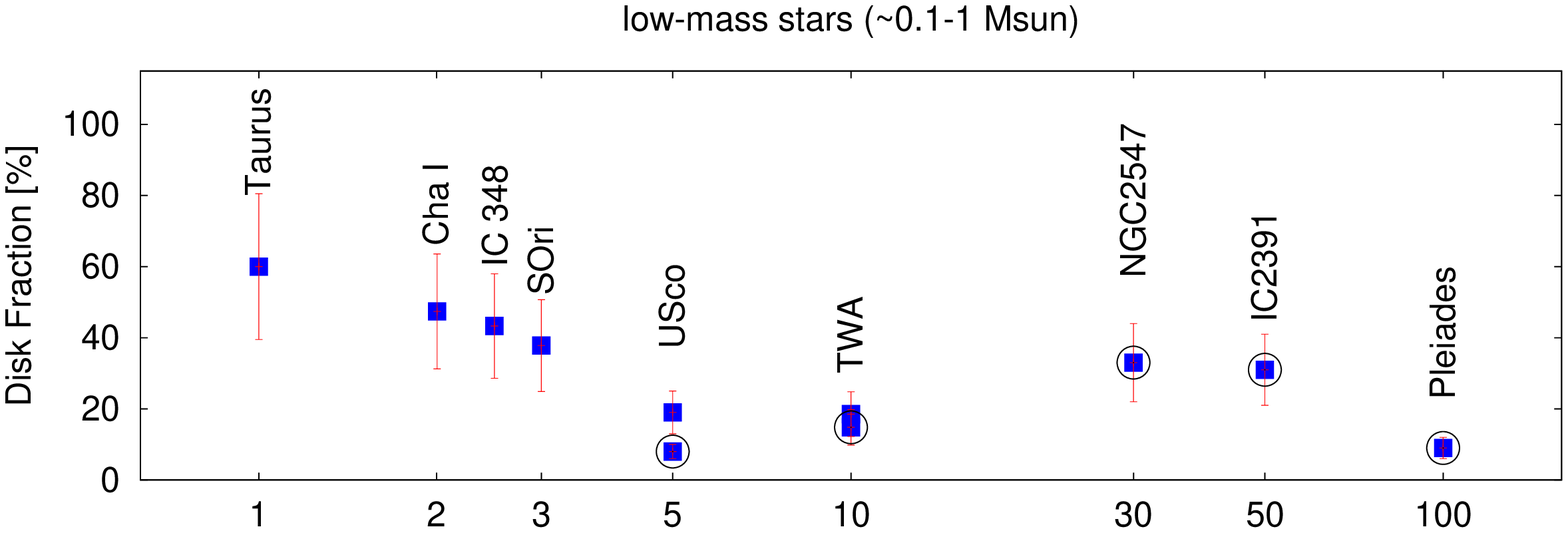}  
  \includegraphics[width=14cm, angle=0]{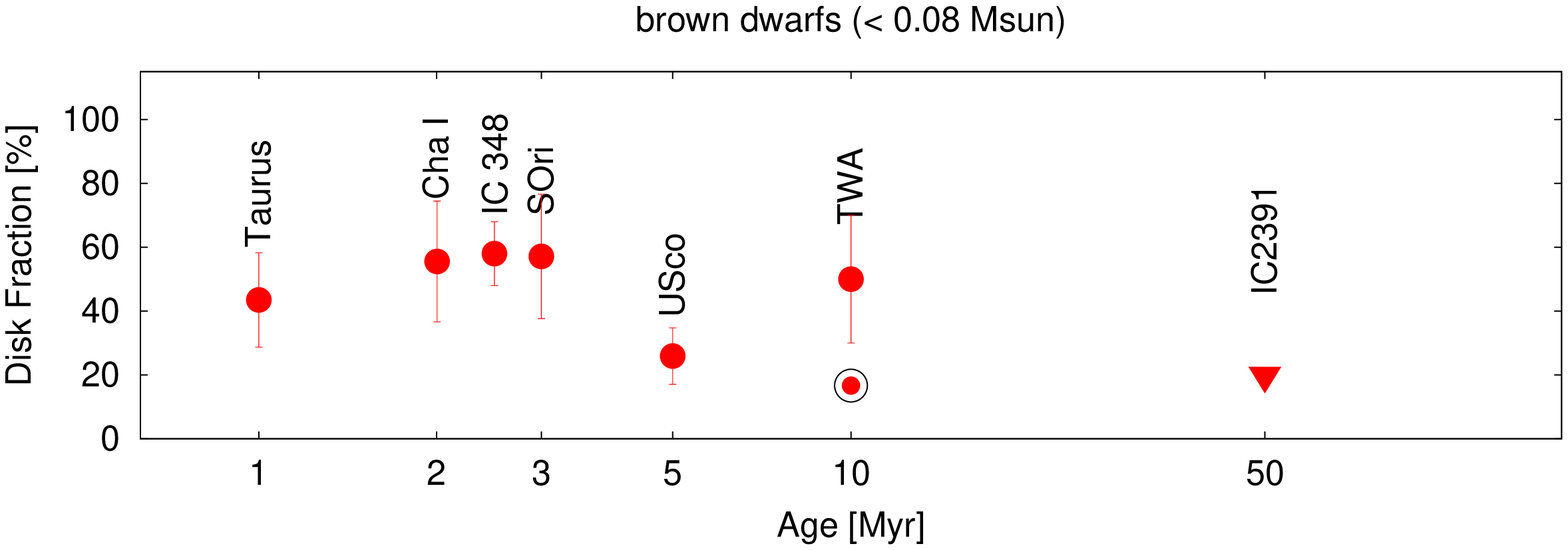}    
 \caption{The disk fractions vs. stellar age for the high-mass stars (top), low-mass stars (middle) and brown dwarfs (bottom). Debris disk fractions are circled in black. The primordial disk fractions are based on the presence of excess emission at $\leq$10$\mu$m wavelengths, whereas the debris disks are photospheric at these wavelengths but show excess at 24 and/or 70$\mu$m. For IC 2391 brown dwarfs, the red triangle denotes the upper limit to the primordial disk fraction.  }
  \label{frac}
\end{figure*}

In Fig.~\ref{frac}, we have plotted the disk frequencies at ages of $\sim$1-100 Myr, for three different mass bins. For the `young' ($\sim$1-10 Myr) group, we have used the spectral type range of B5-K5 for the bin of high-mass stars ($\sim$1-4 $M_{\sun}$), the spectral range K7-M5 for low-mass stars ($\sim$0.1-1 $M_{\sun}$), and spectral type later than M5 for brown dwarfs ($<$0.08$M_{\sun}$). The (rough) thresholds for the bins have been determined using the evolutionary models by Baraffe et al. (2003). For the `old' age group ($>$10 Myr), given the discrepancies in the spectral type coverage from the various surveys, it is difficult to use the same spectral type boundaries for all clusters. At ages of $\sim$30-100 Myr, the $\sim$1 $M_{\sun}$ mass boundary would roughly correspond to a K0-K3 spectral type. However, most surveys at these ages have focused on BAFG stars, with a few, if any, KM-type stars included in the sample. To be consistent with previous comparative studies on debris disks, we have used the same spectral type range in a bin for these older clusters as considered by Siegler et al. (2007). In NGC 2547, the high-mass bin consists of B8-A9 stars, while the low-mass bin covers the F0-F9 spectral range. In IC 2391, the high mass bin has B5-A9 stars, and low-mass covers the F2-K4 range. In Pleiades, the high-mass range is the same as IC 2391, while the low-mass range has F3-K6 stars.
 
The data points in Fig.~\ref{frac} circled in black are debris disk fractions, while the rest of the points represent primordial disk fractions. The criteria for distinguishing between primordial and debris disks is described in \citet{c09} and \citet{r12}. The primordial disks, as they are defined, are the ones that show excess emission at $\leq$10$\mu$m wavelengths, which probe the inner regions at radii within $\sim$1--5 AU in these disks. Debris disks are photospheric at these wavelengths, but show excess emission at 24$\mu$m and/or 70$\mu$m, which arises from the outer disk regions at radii $>$1--5 AU. A compilation of the disk fractions with references is provided in Table~\ref{diskfrac}. We have also compared our Upper Scorpius (USco) fractions with the ones provided for different spectral type bins in Luhman \& Mamajek (2012). These bins have slightly different boundaries than the ones we have considered. The criteria used for classifying primordial/debris disks are also different. If we combine the Luhman \& Mamajek fractions for their B8-M0 bins, then the primordial fraction (from 8$\mu$m excess) is $\sim$7\%, and the debris disk fraction (from 24$\mu$m excess) is $\sim$25\%. These fractions are higher but still comparable within the uncertainties to the USco fractions for the high-mass bin in Table~\ref{diskfrac}. Likewise, the low-mass and brown dwarf bin fractions for USco in Table~\ref{diskfrac} are comparable but slightly higher than the M0-M4 and M4-M8 fractions from Luhman \& Mamajek (2012). These slight differences in fractions are expected due to the different spectral type bins. We note that a recent analysis by Pecaut et al. (2012) indicates the possibility for the age of USco to be older ($\sim$11 Myr) than the widely considered estimate of $\sim$5 Myr in various disk studies (e.g., Carpenter et al. 2009; Riaz et al. 2012). Pecaut et al. have argued that an older age would be consistent with a null excess fraction for F-type stars in USco, with the excess frequency determined from H$\alpha$ and/or $K$-band excess emission. However, these indicators probe the warm inner edge of the disk, and the disk material in the innermost ($<$0.1 AU) region is observed to be dissipated within 2-3 Myr of age (e.g., Haisch et al. 2001). If USco is indeed as old as the TWA, then we would not expect to see excess emission in the mid-infrared IRAC or WISE bands at wavelengths shortward of $\sim$10-12$\mu$m, which probe the primordial disk material at radii within $\sim$1--2 AU. For all three categories of stars shown in Fig.~\ref{frac}, an older age for USco would not be consistent with the still existent primordial disks in this association, based on the mid-infrared excess emission. We have therefore placed this point at 5 Myr, following previous disk studies. 

By an age of $\sim$5--10 Myr, the primordial disk fractions for the high-mass and low-mass stars have reached their lowest point; only $\sim$10--20\% of the disks at these ages are in the primordial phase, which indicates that a large number of these disks have experienced significant inner disk clearing \citep[e.g.,][]{c09}. The disk fractions then show a rise again, from $\sim$20\% at 10 Myr to $\sim$40\% at $\sim$30 Myr. These older disks are found to be optically thin and gas-poor, and have been classified as debris disk systems \citep[e.g.,][]{rieke, siegler}. Thus by an age of $\sim$10 Myr, the inner disk material is significantly dissipated for the high- and low-mass stars, and the disks have made a clear transition to the debris phase. The rise in the debris disk fraction for the high-mass stars between $\sim$50 and 100 Myr could be a sample size difference, since the Pleiades sample studied by Gorlova et al. (2006) is about twice the BA-type star sample studied by Siegler et al. (2007) in IC 2391. 

The bottom panel in Fig.~\ref{frac} shows the disk fractions for brown dwarfs, which are based on $\sim$3--24$\mu$m observations, and probe radii $<$1 AU in these disks. No clear age dependence is evident in the case of the sub-stellar sources, as the primordial disk fraction appears to be nearly constant between 1 and 10 Myr, except a dip at 5 Myr, which is likely due to a different formation mechanism for Upper Scorpius brown dwarfs and/or the higher brown dwarf to star number ratio in this particular association, as argued in \citet{r12}. The general picture thus indicates that protoplanetary disks around brown dwarfs tend to remain in the primordial stage for a relatively longer timescale compared to higher mass stars \citep[e.g.,][]{rg08}. As mentioned, none of the eight IC 2391 targets were detected in the WISE 12 and 22$\mu$m bands. The WISE 5-$\sigma$ point source sensitivity at 12 and 22$\mu$m is $\sim$0.7 and $\sim$5 mJy, respectively (WISE Explanatory Supplement). The predicted photospheric fluxes at these wavelengths for an M6 dwarf are $\sim$0.1--0.2 mJy. Thus while it is impossible to detect the photospheres at these wavelengths, the IC 2391 brown dwarfs may possess weak mid-infrared excesses, with an observed to photospheric flux ratio $F_{obs}$/$F_{phot}$ of $<$3. In comparison, the primordial brown dwarf disks in the $\sim$10 Myr old TWA exhibit large 12$\mu$m excesses, with $F_{obs}$/$F_{phot}$ $>$8--10 \citep{r12}. The non-detections for the IC 2391 cases then suggests that there has been significant inner disk clearing, and that the disks around brown dwarfs have transitioned to the debris phase by $\sim$40--50 Myr ages, resulting in photospheric emission and/or negligible excesses at mid-infrared wavelengths. 

A similar argument can be applied for the PACS wavebands. If we compare the PACS 70$\mu$m upper limits for the IC 2391 targets with the observed flux densities in this band for the primordial 1--10 Myr brown dwarf disks \citep{rg12, harvey}, then nearly 80\% of these disks lie above the 2-$\sigma$ upper limit, while the rest have fluxes at the 1-$\sigma$ limit. Thus there is a strong likelihood that any primordial disks with similar brightness and fractional luminosities as observed for the younger brown dwarfs would be within our survey detection limits and could have been detected. We can test this by applying a simple exact binomial test using the null hypothesis that there is an equal probability of detecting or not detecting a primordial disk in a single observation. The one-tailed probability of detecting a primordial disk for 6 out of 8 IC 2391 targets is quite high, about 96\%. The P-value of 0.96 is significantly strong, being nearly a factor of 20 higher than the significance level of 0.05, thus making the null hypothesis valid. The null detection rate despite such a high probability again indicates that a majority of the brown dwarf disks have transitioned to the debris stage by this age. Given that there is presumably a range of primordial disk brightnesses, the fraction of primordial brown dwarf disks for IC 2391 cannot be definitively set to 0\%, but can perhaps be set to $<$20\% (Fig.~\ref{frac}), in comparison with the detection likelihood of the younger disk sources. We note that a large number of primordial brown dwarf disks have not been observed yet at PACS wavelengths, and so the fraction that lies above our detection limits could be smaller than $\sim$80\%. Nevertheless, our PACS survey has helped set limits on the primordial disk lifetime to be $<$40 Myr for the sub-stellar sources.

The PACS observations, unfortunately, do not provide any constraints on the debris disk evolution around brown dwarfs. The conclusion from Fig.~\ref{lum} was that typical debris disks with fractional luminosities similar to those observed for low-mass stars could {\it not} have been detected given the sensitivities of our survey. In other words, Vega-like debris disks at $L_{D}$/$L_{*}$ $\leq$ 10$^{-3}$ levels could exist around the IC 2391 brown dwarfs, and the fraction could be as high as $\sim$30\%, as observed for low-mass stars at these ages (e.g., Siegler et al. 2007). It is worth noting that there is one candidate brown dwarf debris disk in the TWA (2MASS J1139511-315921), which shows a small excess at 24$\mu$m with $F_{obs}$/$F_{phot}$$\sim$3, indicating the presence of warm dust around it. This object was undetected in the follow-up PACS 70$\mu$m observations, and it was argued that the debris disk has a very low fractional luminosity of $L_{D}/L_{*}$$\sim$10$^{-5}$, thus lying below the detection limit \citep{rg12}. The presence of this candidate suggests a transition to the debris phase for brown dwarfs may have begun at $\sim$10 Myr, and perhaps all sub-stellar debris systems are at such low fractional luminosities of $\sim$10$^{-5}$. The debris disk fraction might increase thereafter between 10 and 40 Myr, or it might rapidly decline during this short timescale. A rapid debris dispersal would be consistent with the trends noted by e.g., \citet{gorl} and \citet{plavchan} for M-type debris systems, and by e.g., \citet{siegler} for the FGK debris disks, which appear to decay over a comparatively faster timescale than BA type stars, although these trends are not definitively confirmed, and the samples studied for M-type stars are much smaller in size compared to the earlier type sources. 


\begin{figure*}
\centering
  \includegraphics[width=8.5cm, angle=0]{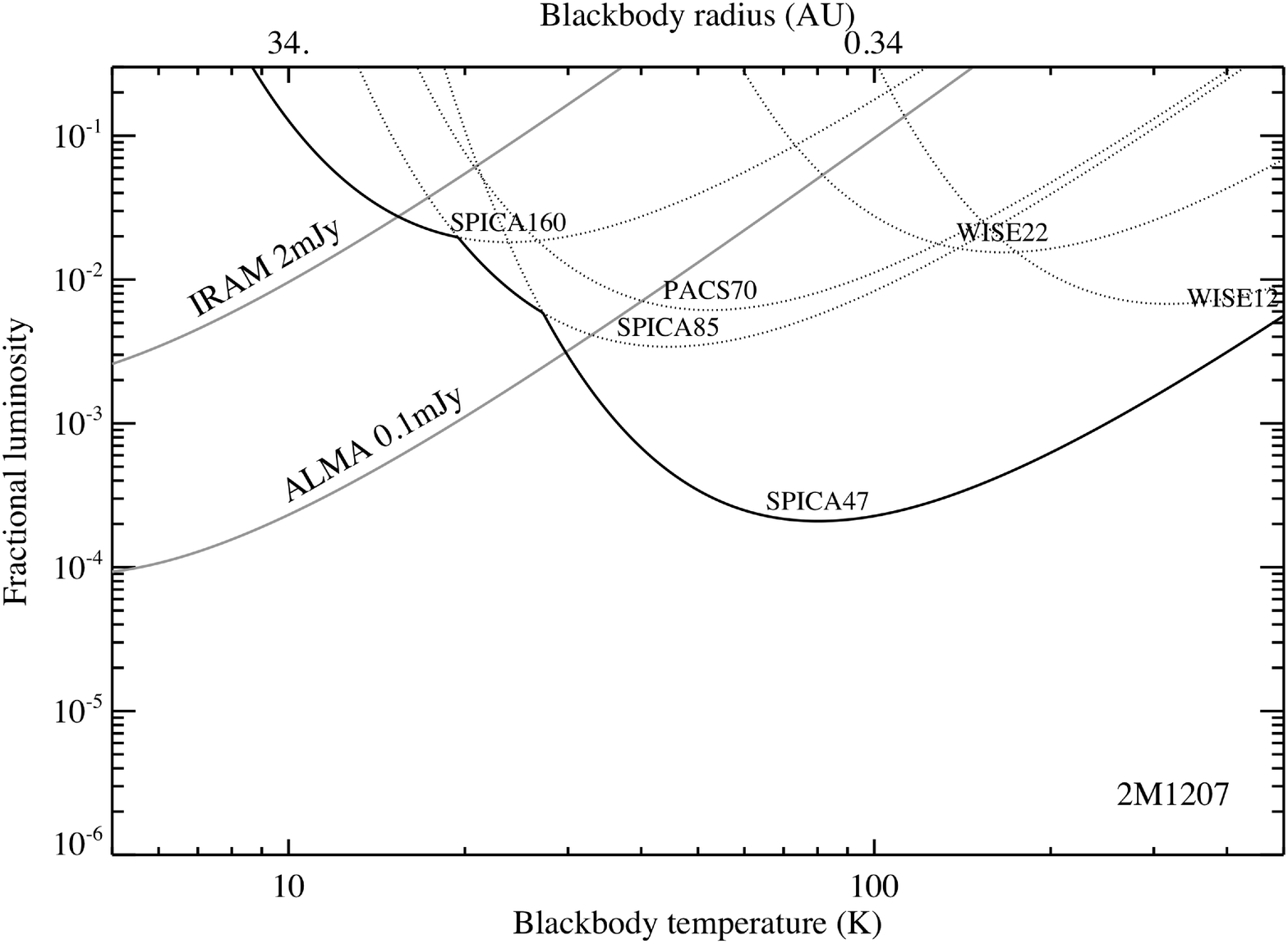} 
 \caption{A sensitivity plot for the $\sim$10 Myr old 2M1207 brown dwarf disk in the TWA. Axes represent the same parameters as shown in Fig.~\ref{lum}. Further details are provided in the text.  }
  \label{lum2}
\end{figure*}

Future observations particularly with the SAFARI imager onboard the SPICA space observatory, will be ideal to complete the census of brown dwarf disks, both in the primordial and debris phases. As an example, in Fig.~\ref{lum2}, we have plotted the sensitivities similar to Fig.~\ref{lum} for the case of the $\sim$10 Myr old brown dwarf disk 2MASSW J1207334-393254 (2M1207) in the TWA, located at a distance of $\sim$50 pc (Mamajek 2005). This is presently the oldest known brown dwarf disk which has been detected in the far-infrared. The WISE and PACS limits are from actual observations (Harvey et al. 2012). The SPICA sensitivities have been calculated using the confusion limits of 0.015, 0.5, and 5 mJy in the SW (34-60$\mu$m), MW (60-110$\mu$m), and LW (110-210$\mu$m) bands. The IRAM and ALMA limits are as marked. The main conclusion that can be made from this figure is that for probing dust temperatures of a few tens to a few hundreds of Kelvins at distances of within 1 AU or the habitable zone from a central brown dwarf, the SPICA SW band would be the most suitable one, simply because the confusion limit is much fainter. ALMA is better at cooler temperatures. The IRAM/NIKA 1.2mm band, however, would be sensitive to brighter disks of $>$1\% fractional luminosity, similar to the sensitivity of our present PACS survey. We also note that the 2M1207 disk was undetected in the SPIRE $\sim$200-500$\mu$m observations, which were sensitive to $\sim$2 mJy level (Riaz \& Gizis 2012). Therefore, high-quality observations of large-sized samples with SPICA and ALMA in the far-infrared/sub-millimeter domain should provide clues to the dispersal processes and timescales for disks around sub-stellar objects.

\section{Summary}

We have conducted a search for disks around brown dwarfs in the $\sim$40--50 Myr cluster IC 2391. None of the 8 targets were detected in any of the PACS bands. We estimate our survey to be sensitive to fractional debris disk luminosities of $>$1\%, and therefore only the brightest debris disk systems, if they existed, could have been detected. We suggest that debris disks of similar levels ($L_{D}$/$L_{*}$$\leq$10$^{-3}$) as those observed around low-mass stars at $\sim$40--50 Myr ages could exist around the IC 2391 targeted brown dwarfs, but be missed by our survey. It may also be the case that disk dispersal processes such as grain growth and terrestrial planet formation occur at a faster rate around brown dwarfs, once they have entered the secondary debris phase, resulting in a non-detection at later ages. Most primordial disks with flux densities similar to those observed among younger $\sim$1--10 Myr brown dwarfs would be within our survey detection limits and could have been detected. The upper limits from our PACS survey can thus set a constraint of $<$40 Myr on the primordial disk lifetime for brown dwarfs. 

\section*{Acknowledgments}

This work was supported by the European Union through ERC grant number 279973 (GMK). {\it Herschel} is an ESA space observatory with science instruments provided by European-led Principal Investigator consortia and with important participation from NASA. This publication makes use of data products from the Wide-field Infrared Survey Explorer, which is a joint project of the University of California, Los Angeles, and the Jet Propulsion Laboratory/California Institute of Technology, funded by the National Aeronautics and Space Administration.

\begin{onecolumn}
\begin{landscape}
\begin{table*}
\begin{minipage}{\linewidth}
\small
\caption{Observations}
\label{phot}
\begin{tabular}{@{}c|c|c|@{}c|@{}c|@{}c|@{}c|@{}c|@{}c|@{}c|@{}c|@{}c|@{}c|@{}c|@{}c|@{}c|@{}c|@{}c|@{}c}
\hline

ID &  Position (J2000) & ObsID   &     T$_{eff}$ & log L & SpT &  3.4$\mu$m\footnote{WISE photometry. A `null' error and a negative SNR imply a non-detection.} (err) &    SNR  &   4.6$\mu$m (err) &    SNR  &   12$\mu$m (err)   &  SNR  &   22$\mu$m (err)  &   SNR & 70$\mu$m\footnote{PACS 70 and 160$\mu$m 1-$\sigma$ flux upper limits in units of mJy.}  & 160$\mu$m  \\   \hline
1 & 083847.07-521456.4 & 1342265624/5 & 2800 & -2.85 & M6 & 14.00 (0.025)&  43.7 &13.726 (0.029) & 37.6 &	12.479 (null) &	-0.6 &	9.322 (null) &	-1.3    & 0.93 & 5.26 \\
2 & 083847.30-524432.7 & 1342265620/1 & 2800 & -2.575 & M6 & 13.864 (0.026)&  41.6 &13.589 (0.029)&	37.7 & 13.176 (null) & -2.3 &	9.336 (null) &	-2.1     & 1.54 & 17.38 \\
3 &  084218.71-523940.1 & 1342265505/6 & 2575 &  -2.70 & M7 & 13.569 (0.024)&  44.6&13.408 (0.026)&	41.3&	12.638 (null) &	-0.7&	9.313 (null) &	-0.5	& 0.92 & 7.44 \\
4 &  084323.67-531416.9 & 1342265503/4 & 2800 & -2.75 & M6 &  13.762 (0.025)&  42.9&13.690 (0.028)& 39.3&   12.259 (null) & 0.6&   9.095 (null) &  0.2    & 0.89 & 6.66 \\
5\footnote{The photometry for the target ID \# 5 was found in the AllWISE Reject Table, because it is affected by a diffraction spike from a nearby star.} & 084402.10-524410.7 & 1342270971/2 & 2575  & -2.70 & M7 & 13.612 (0.034) & 32.3    & 13.432 (0.035)  & 31.2  &  12.088 (0.22) & 4.9  &  9.172 (null) & -0.6 & 1.15 & 14.72 \\
6 & 083618.24-532557.6 & 1342270969/70 & 2654 & -2.89 & M7.5 & 14.059 (0.026)&  41.1 &13.831 (0.033)&	33.2 &	12.966 (null) &	-0.2&	9.514 (null) &	-1.9	& 0.85 & 3.46 \\
7 & 083933.26-524710.4 & 1342265507/8 & 2958  & -2.55 & M6.5    & 13.639 (0.030)     & 35.6   &  13.416 (0.030)   &   36.0   &  12.372 (0.236)   & 4.6   & 8.962 (null)   &  1.1    & 1.15 & 7.38 \\                                                                              
8 & 084016.67-523658.3 & 1342265622/3 & 2740 & -2.81 & M6.5 & 14.616 (0.027)& 40.8&14.424 (0.044)&	24.5&	12.790 (null) &	-0.9&	8.978 (null) &	0.3	& 0.98 & 7.26 \\

\hline
\end{tabular}
\end{minipage}
\end{table*}
\end{landscape}
\end{onecolumn}

\begin{onecolumn}
\begin{landscape}
\begin{table*}
\begin{minipage}{\linewidth}
\caption{Disk Fractions}
\label{diskfrac}
\begin{tabular}{c|c|c|c|c}
\hline

Name  &  \multicolumn{3}{c}{Disk Fraction [\%]}  & References \\  \hline
             & high-mass & low-mass & brown dwarfs  &  \\
\hline

Taurus$^{a}$ & 75$\pm$26 & 60$\pm$20 & 43$\pm$15  & 1, 2  \\
Cha I$^{a}$ & 65$\pm$22 & 47$\pm$16 & 55$\pm$19 & 1, 2 \\
IC 348$^{a}$ & 42$\pm$14 & 43$\pm$15 & 58$\pm$10 & 1, 2 \\
$\sigma$ Orionis$^{a}$ & 28$\pm$10 & 38$\pm$13 & 57$\pm$19 & 1, 2 \\
USco$^{b}$ & 4$\pm$3, 18$\pm$6 & 19$\pm$6, 8$\pm$2 & 26$\pm$9 & 3, 4, 5, 6 \\
TWA$^{c}$ & 50$\pm$20 & 9$\pm$3, 18$\pm$6 & 50$\pm$20 & 4, 7, 8 \\
NGC 2547$^{d}$ & 44$\pm$15 & 33$\pm$11 & -- & 11, 12 \\
IC 2391$^{e}$ & 10$\pm$3 & 31$\pm$10 & $\leq$30 & 13, 14 \\
Pleiades$^{f}$ & 25$\pm$8 & 9$\pm$3 & -- & 15, 16 \\
                     
\hline
\end{tabular}

{\it a:} These discs are classified as primordial/evolved disks. The debris disk fraction for the younger clusters is negligible ($\sim$1-2\%). 

{\it b:} The first value is for primordial disks, second value for debris disks. The brown dwarf disks are all classified as primordial/evolved disks. The total disk fraction for high-mass stars is 22$\pm$7\%, and for low-mass stars is 27$\pm$9\%. 

{\it c:} For the high-mass stars, all disks are classified as debris sources. For the low-mass stars, the first value is for primordial disks, second value for debris disks. The brown dwarfs are all classified as primordial/evolved disks. The total disk fraction for low-mass stars is 21$\pm$7\%.

{\it d:} High-mass and low-mass bins cover the spectral ranges of B8-A9 and F0-F9, respectively. 

{\it e:} High-mass and low-mass bins cover the spectral ranges of B5-A9 and F2-K4, respectively. 

{\it f:} High-mass and low-mass bins cover the spectral ranges of B5-A9 and F3-K6, respectively. 

REFERENCES: (1) Luhman et al. (2008); (2) Luhman et al. (2010); (3) Carpenter et al. (2009); (4) Riaz et al. (2009); (5) Scholz et al. (2007); (6) Riaz et al. (2012); (7) Low et al. (2005); (8) Schneider et al. (2012); (9) Su et al. (2006); (10) Chen et al. (2005); (11) Gorlova et al. (2007); (12) Young et al. (2004); (13) Siegler et al. (2007); (14) This work; (15) Gorlova et al. (2006); (16) Stauffer et al. (2005).

\end{minipage}
\end{table*}
\end{landscape}
\end{onecolumn}

\label{lastpage}

\end{document}